\documentclass[runningheads]{llncs}
\usepackage[T1]{fontenc}
\usepackage{graphicx}
\usepackage{amsmath,amssymb}
\usepackage{microtype}
\usepackage{flafter}
\usepackage[section]{placeins}
\usepackage{xurl}
\usepackage[hidelinks]{hyperref}
\graphicspath{{figures/}}
\begin{document}
\title{Beyond Gate Weights: Decision-Level Interpretation of Radiomics--Image Feature Fusion for RCC Classification}
\titlerunning{Interpreting Gated Radiomics--Image Fusion}
\author{
Yuan Liang\inst{1,3} \and % index{Liang, Yuan}
Sourav Bhattacharjee\inst{2} \and % index{Bhattacharjee, Sourav}
Abraham Campbell\inst{3}  % index{Campbell, Abey}
}
\authorrunning{Y. Liang et al.}
\institute{
Research Ireland Centre for Research Training in Machine Learning  
\and
School of Veterinary Medicine, University College Dublin, Dublin, Ireland \and
School of Computer Science, University College Dublin, Dublin, Ireland 
\email{yuan.liang@ucdconnect.ie} \\
}
\maketitle

\begin{abstract}
Gate coefficients describe how a fusion layer weights intermediate features, but they do not necessarily reflect the magnitude of branch contributions at the classifier output. We investigate this distinction in a fixed 3D MedVAE--radiomics model for clear cell versus non-clear cell renal cell carcinoma classification from contrast-enhanced CT. Because the model uses a linear classification head, the binary decision logit can be decomposed exactly into a gated image term, a direct radiomics term, and a bias. We define effective radiomics dominance as the direct radiomics term's share of the summed absolute non-bias branch contributions and compare it with structural radiomics weighting. Internally, mean structural radiomics weighting was 0.823, whereas mean effective radiomics dominance was 0.148, with a modest case-wise association ($\rho=0.290$). The gated image term retained substantially stronger discriminative ranking information than the direct radiomics term. The same structural-to-effective discrepancy was observed in exploratory external evaluation. Branch-targeted permutation, integrated-gradient, and Shapley-style analyses further identified radiomics descriptors influencing the direct radiomics term. Because radiomics also generates the gate, the gated image term remains radiomics-dependent, so additive branch contribution should not be interpreted as total input influence. These findings show that structural gate weighting and decision-level branch contribution represent distinct aspects of gated feature fusion and should be interpreted separately.
\end{abstract}

\section{Introduction}

Renal cell carcinoma (RCC) is a major urological malignancy for which preoperative subtype characterisation is clinically relevant. Clear cell RCC (ccRCC) and non-clear cell RCC differ in prognosis and management, while their appearances can overlap on contrast-enhanced computed tomography (CT) \cite{bex2025european}. Radiomics describes tumour shape, intensity distribution, and texture through predefined measurements \cite{aerts2014decoding}. CT radiomics has been investigated for distinguishing ccRCC from non-clear cell RCC \cite{wang2021radiomics}. Learned image representations offer another description of tumour appearance, including transferable latent features from medical image encoders such as MedVAE \cite{pmlr-v301-varma26a}. 

Gated fusion adaptively weights intermediate feature branches \cite{arevalo2017gated}. Gate coefficients have also been used to interpret modality emphasis in clinical prediction. For example, Sun et al. analysed softmax self-gating weights across electronic-health-record time windows and complemented this analysis with integrated-gradient attribution \cite{sun2025borrowing}. Such approaches motivate a distinction between weighting at the fusion layer and contribution to the final decision. The latter additionally depends on feature magnitudes, their directions, and downstream classifier weights. When radiomics generates the gate, it influences both its own branch and the weighting of the image branch, so direct additive contribution and total input dependence are distinct quantities.

The interpretation of attention coefficients has been debated in relation to other measures of model behaviour \cite{jain2019attention,wiegreffe2019attention}. Kobayashi et al. analysed transformed-vector norms alongside attention weights, obtaining a different account of information aggregation in Transformers \cite{kobayashi2020attention}. Although attention and the gated classifier studied here have different architectures, this work motivates examining weighted features and their downstream effects. In medical imaging, specifying the model quantity being interpreted helps establish the relevance of an analysis to clinical questions \cite{borys2023explainable}.

We study a fixed 3D MedVAE--radiomics gated classifier through three complementary analyses: decomposition of its decision logit, comparison of structural and effective branch weighting, and feature attribution targeted to its direct radiomics term. Internal and exploratory external evaluation assess discrimination, directional agreement, and radiomics feature sensitivity. Together, these analyses connect gate behaviour to signed classifier-level contributions and provide a focused investigation of radiomics--image fusion in RCC.

\section{Methods}

\subsection{Cohorts and Image Preprocessing}

The internal cohort comprised 396 patients from KiTS23, a collection of contrast-enhanced abdominal CT scans with kidney and renal tumour annotations \cite{heller2023kits23}. Multifocal cases were excluded to avoid ambiguity between patient histology and the analysed lesion. The binary endpoint was ccRCC ($Y=1$) versus non-clear cell RCC ($Y=0$). A fixed stratified patient-level split assigned 237 patients to training (168 ccRCC, 69 non-clear cell RCC), 99 to validation (70 and 29), and 60 to internal testing (42 and 18). Classifier parameters were fitted on training data, and validation data were used for checkpoint and classification-threshold selection. The held-out test partition was used only for final evaluation and was not used for model fitting.

CT images and masks were reoriented to RAS and resampled to 1.0-mm isotropic spacing, using linear interpolation for images and nearest-neighbour interpolation for masks. Intensities were clipped to $[-150,200]$ HU. The learned image branch used tumour-centred crops with a 12-voxel contextual margin. Nonzero image voxels were z-score normalised per volume, followed by padding or centre cropping to $96\times96\times96$ voxels. Radiomics extraction used HU-scale images without this image z-score normalisation. For radiomics, multiple annotations of the same lesion were combined by STAPLE \cite{warfield2004simultaneous}.

The external TCGA/AIMI cohort comprised 41 cases, including 39 ccRCC and two non-clear cell RCC cases. Cases were drawn from the TCGA-KIRC, TCGA-KIRP, and TCGA-KICH collections, using AIMI DICOM-SEG annotations \cite{murugesan2024ai,vanoss2023aimi}. Cohort construction retained expert-corrected masks or masks with the highest available expert quality score. External cases were not used to fit the classifier, select radiomics features, estimate training normalisation statistics, or select the checkpoint or classification threshold. The external cohort was used for exploratory evaluation of the fixed model.

\subsection{Radiomics Representation}

PyRadiomics was used to extract shape, first-order, and texture features \cite{van2017computational}, with reference to IBSI feature definitions \cite{zwanenburg2020image}. Shape features were computed on the original tumour region. First-order and texture features were computed on original and Laplacian-of-Gaussian (LoG) images with $\sigma\in\{1,2,3\}$~mm, using a fixed discretisation bin width of 25 HU. The initial set contained 386 features.

The inherited pipeline first required inter-annotator ICC(3,1), a two-way mixed-effects consistency measure \cite{shrout1979intraclass}, of at least 0.75. Features below the 20th percentile of median absolute deviation (MAD) were then removed, followed by greedy Pearson-correlation filtering at $|r|>0.95$. Average-linkage hierarchical clustering used distance $1-|r|$ and a cut threshold of 0.2, retaining the feature with the largest MAD from each cluster. The classifier analysed here used the resulting fixed 73-feature representation. Feature-wise z-score statistics were estimated from the 237 training patients and applied unchanged to the validation, internal test, and external cohorts.

\subsection{Gated Classifier and Model Development}

The image branch used the volumetric MedVAE encoder \cite{pmlr-v301-varma26a}. Its latent tensor was flattened and linearly projected to $\mathbf f_{\mathrm{img}}\in\mathbb R^{512}$. The 73-feature radiomics vector was independently projected to $\mathbf f_{\mathrm{rad}}\in\mathbb R^{512}$. A radiomics-conditioned multilayer perceptron, with dimensions $512\rightarrow256\rightarrow512$, a hidden ReLU, and an output sigmoid, produced the feature-wise image gate:
\begin{equation}
\mathbf g=\sigma\!\left(W_2\operatorname{ReLU}(W_1\mathbf f_{\mathrm{rad}}+\mathbf a_1)+\mathbf a_2\right).
\end{equation}
The fused representation and two-class linear head were
\begin{equation}
\mathbf h=\mathbf g\odot\mathbf f_{\mathrm{img}}+(1-\mathbf g)\odot\mathbf f_{\mathrm{rad}},
\qquad (z_0,z_1)^\top=W_{\mathrm{cls}}\mathbf h+\mathbf b_{\mathrm{cls}},
\label{eq:gatedfusion}
\end{equation}
where $\odot$ denotes element-wise multiplication. A projected dimension is a learned mixture of input features, not an individually identifiable radiomics descriptor.

Model development used partial fine-tuning, class-weighted cross-entropy, Adam, and a batch size of 2. Training augmentation comprised independent flips along two spatial axes (probability 0.5 each), random 90-degree rotations (0.2), and affine rotations/scaling (0.2; rotation range 0.1 radians and scale range 0.1 per axis). Validation and evaluation preprocessing were deterministic. The learning rate was halved after five epochs without improved validation AUC; early stopping used patience 15, and the checkpoint with the highest validation AUC was retained. All contribution analyses used this fixed checkpoint and its unscaled scores.
% AUTHOR CHECK -- historical training provenance:
% The supplied draft's backbone/head rates 1e-5/1e-4 and named-stage policy
% match a later script, whereas recovered earlier source uses single-group
% Adam at 1e-4 and last-fraction unfreezing. Neither source establishes the
% exact launch configuration of partial03. Do not insert either as verified
% history without a checkpoint-linked run record. See REVISION_NOTES_ZH.md.

\subsection{Decision-Level Branch Decomposition}

Because the fused representation is followed by a linear classification head, the binary decision logit can be decomposed exactly into additive terms associated with the two gated branches. Let $\mathbf{w}=\mathbf{W}_{\mathrm{cls},1}-\mathbf{W}_{\mathrm{cls},0}$ and $b=b_{\mathrm{cls},1}-b_{\mathrm{cls},0}$. The ccRCC versus non-clear cell RCC logit difference is
\begin{equation}
z=z_1-z_0=\mathbf{w}^{\top}\mathbf{h}+b,
\qquad
\hat{p}=\sigma(z).
\end{equation}
Substituting the gated representation from Eq.~\eqref{eq:gatedfusion} gives
\begin{equation}
C_{\mathrm{img}}=\mathbf{w}^{\top}(\mathbf{g}\odot\mathbf{f}_{\mathrm{img}}),
\qquad
C_{\mathrm{rad}}=\mathbf{w}^{\top}((1-\mathbf{g})\odot\mathbf{f}_{\mathrm{rad}}),
\end{equation}
such that
\begin{equation}
z=C_{\mathrm{img}}+C_{\mathrm{rad}}+b.
\label{eq:decomposition}
\end{equation}
We refer to $C_{\mathrm{img}}$ as the gated image term and $C_{\mathrm{rad}}$ as the direct radiomics term. Both are conditional on the radiomics-generated gate, so this is a decomposition into computational branch terms rather than an allocation of total input influence. For the saved-output analysis, $C_{\mathrm{rad}}$ was reconstructed from the fixed model weights and standardised radiomics inputs, and $C_{\mathrm{img}}$ was recovered as $z-C_{\mathrm{rad}}-b$ from the saved class logits.

For each case, structural radiomics weighting was defined as
\begin{equation}
S_{\mathrm{rad}}
=
1-\frac{1}{d}\sum_{j=1}^{d}g_j,
\qquad d=512.
\label{eq:structural}
\end{equation}
This quantity describes the average structural weighting assigned to radiomics at the fusion layer. Effective radiomics dominance was defined as
\begin{equation}
D_{\mathrm{rad}}
=
\frac{|C_{\mathrm{rad}}|}
{|C_{\mathrm{img}}|+|C_{\mathrm{rad}}|}.
\label{eq:effective}
\end{equation}
$D_{\mathrm{rad}}$ measures the relative magnitude of the direct radiomics term among the two non-bias additive branch terms. Values closer to zero indicate dominance of the gated image term, while values closer to one indicate dominance of the direct radiomics term. Because absolute values are used, contribution direction was analysed separately as described below.

\subsection{Discrimination and Directional Behaviour}

We evaluated AUC and average precision (AP) for $z$, $C_{\mathrm{img}}$, and $C_{\mathrm{rad}}$, with larger values oriented toward ccRCC and no post-hoc score reversal. Spearman correlation quantified the association between $S_{\mathrm{rad}}$ and $D_{\mathrm{rad}}$.

Directional agreement was defined by $C_{\mathrm{img}}C_{\mathrm{rad}}>0$, conflict by $C_{\mathrm{img}}C_{\mathrm{rad}}<0$, and a neutral case by a zero product. These signs describe additions to the ccRCC logit, not separate branch classifications. For error-rate comparisons, the full prediction was classified using a threshold selected by maximising Youden's index on the validation set \cite{youden1950index} and then held fixed across evaluation cohorts. We summarised conflict frequency and compared full-model error rates and $D_{\mathrm{rad}}$ distributions between agreement and conflict groups.

\subsection{Radiomics Feature Attribution}

Feature attribution targeted the direct radiomics term $F(\mathbf{x})=C_{\mathrm{rad}}(\mathbf{x})$, where $\mathbf{x}$ is the standardised 73-feature radiomics vector. Repeated permutation importance was the primary analysis \cite{fisher2019mcr}. Each feature was independently shuffled across patients 200 times, separately within each evaluation cohort, while the radiomics projection and radiomics-conditioned gate were recomputed after each perturbation. Importance was quantified by the mean decrease in AUC and AP of $C_{\mathrm{rad}}$; negative values denote improved discrimination after permutation.

Integrated gradients \cite{sundararajan2017axiomatic} and a Monte Carlo Shapley-style analysis \cite{shapley1953value,castro2009sampling} were complementary methods, computed separately within each cohort. Both used the zero standardised-feature vector, corresponding to training-set feature means, as the reference. Integrated gradients were approximated using 64 equally spaced right-endpoint evaluations along the straight-line path to each patient's feature vector; absolute attributions were then averaged across patients. The Shapley-style analysis sampled 64 random orders of all 73 features. Starting from the reference, each feature was replaced by its patient-specific value, and its marginal change in $C_{\mathrm{rad}}$ was recorded. Absolute marginal changes were averaged over orders and patients. This magnitude summary differs from taking the absolute value of each patient's order-averaged Shapley estimate.

\subsection{Statistical Analysis}

AUC/AP 95\% confidence intervals (CIs) were the 2.5th and 97.5th percentiles from 1,000 class-stratified patient bootstrap replicates \cite{efron1979bootstrap}. Paired differences used identical resampled patients for the compared scores. These intervals condition on the observed class counts and fixed model. Spearman CIs used 1,000 ordinary patient bootstrap replicates. Error-rate comparisons used Fisher's exact test, and $D_{\mathrm{rad}}$ distributions were compared using the two-sided Mann--Whitney test. Comparisons were unadjusted for class composition or other covariates. All tests were exploratory and were not corrected for multiple comparisons.

For feature permutation, the 2.5th--97.5th percentile range of the 200 score decreases summarised variability across random shuffles of the fixed cohort. These ranges quantify permutation variability, whereas patient bootstrap CIs quantify patient-sampling uncertainty conditional on the specified resampling design. Neither incorporates uncertainty from training or feature selection.

\section{Results and Discussion}
\label{sec:results}

\subsection{Discrimination of the Decomposed Scores}

The gated image term retained discrimination close to the full classifier, whereas the radiomics branch term showed substantially weaker ranking performance (Table~\ref{tab:performance}). Internally, paired bootstrap analysis gave a full-minus-image AUC difference of $-0.017$ (95\% CI $-0.042$ to 0.000), compared with 0.308 (95\% CI 0.165--0.468) for the full-minus-radiomics difference. These results indicate that substantially more of the observed discriminative ranking information was carried by the gated image term than by the radiomics branch term, without establishing superiority or equivalence between the gated image term and the complete classifier.

\begin{table}[!htbp]
\centering
\caption{Discrimination of the full score and the gated branch terms. Parentheses contain class-stratified patient bootstrap 95\% CIs, conditional on the fixed model. ccRCC prevalence was 0.700 internally and 0.951 externally.}
\label{tab:performance}
\small
\setlength{\tabcolsep}{4pt}
\begin{tabular}{llcc}
\hline
Cohort & Score & AUC (95\% CI) & AP (95\% CI) \\
\hline
Internal & Full score & 0.827 (0.710--0.923) & 0.922 (0.867--0.967) \\
 & $C_{\mathrm{img}}$ & 0.844 (0.738--0.938) & 0.928 (0.877--0.972) \\
 & $C_{\mathrm{rad}}$ & 0.519 (0.352--0.689) & 0.720 (0.629--0.833) \\
\hline
External & Full score & 0.782 (0.641--0.897) & 0.988 (0.979--0.995) \\
 & $C_{\mathrm{img}}$ & 0.756 (0.615--0.872) & 0.987 (0.978--0.993) \\
 & $C_{\mathrm{rad}}$ & 0.462 (0.000--0.974) & 0.926 (0.862--0.999) \\
\hline
\end{tabular}
\end{table}

The same qualitative pattern was observed externally, although the estimates were exploratory given the positive-enriched cohort (Table~\ref{tab:performance}).

The discrimination of $C_{\mathrm{img}}$ should not be interpreted as the performance of an independently trained image-only model. The image representation was learned within the fusion model, and its weighting remains conditioned on radiomics through the gate. The branch-score analysis therefore characterises additive terms within the intact classifier rather than the effect of removing either feature branch.

\FloatBarrier

\subsection{Structural Weighting versus Effective Contribution}

\begin{figure}[!t]
\centering
\includegraphics[width=\textwidth]{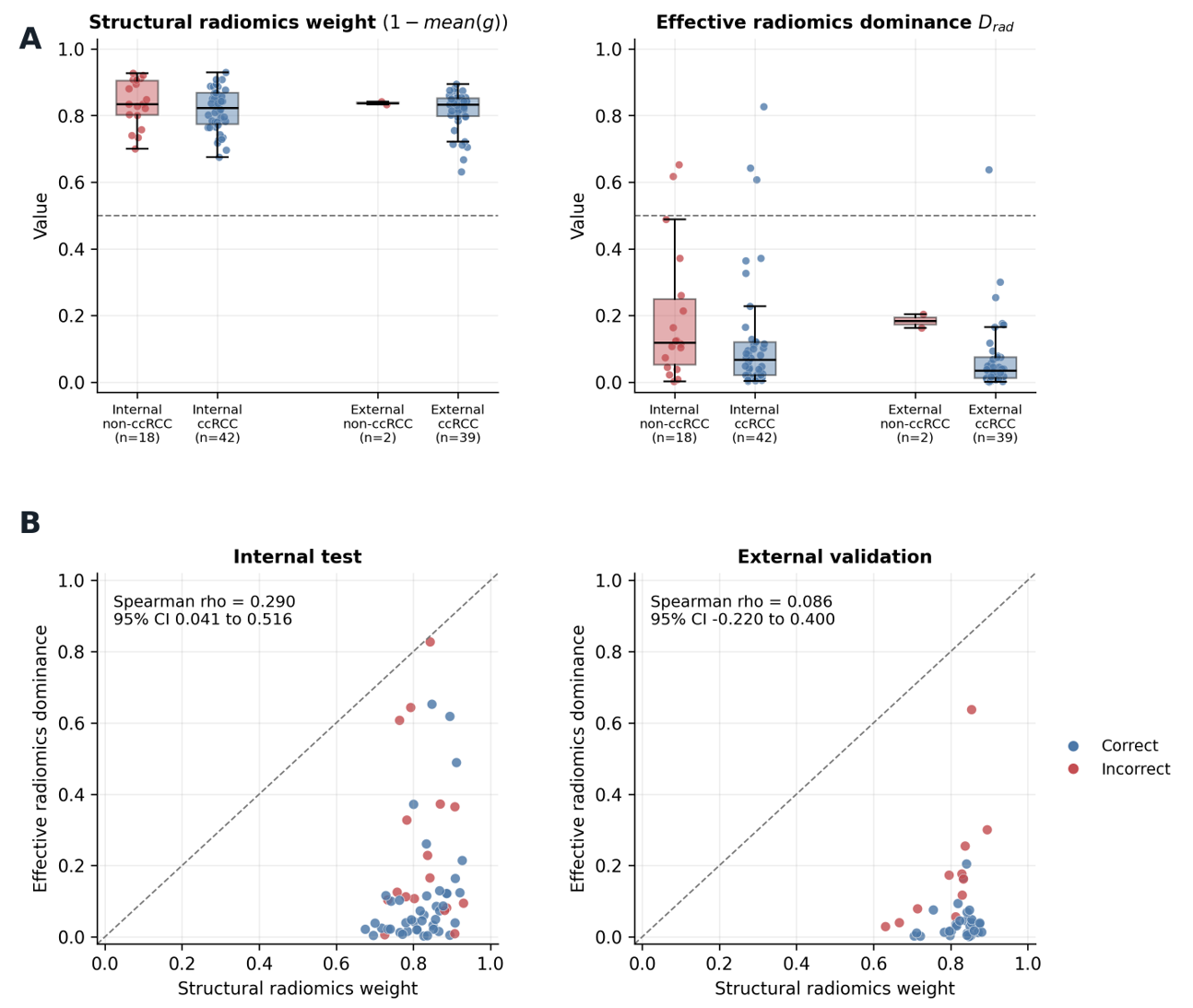}
\caption{Structural and effective radiomics weighting. (A) Distributions by cohort and true class. (B) Case-wise association, coloured by full-model correctness at the validation-selected threshold. Spearman 95\% CIs use ordinary patient bootstrap resampling. The diagonal denotes numerical equality. Both branch terms retain radiomics-conditioned gating.}
\label{fig:structural_effective}
\end{figure}

Mean structural radiomics weighting $S_{\mathrm{rad}}$ was 0.823 internally, whereas mean effective radiomics dominance $D_{\mathrm{rad}}$ was 0.148. The corresponding external means were 0.815 and 0.078 (Fig.~\ref{fig:structural_effective}A). Thus, strong structural weighting toward radiomics coexisted with a substantially smaller radiomics share of the absolute net additive branch terms.

Structural radiomics weighting showed a modest rank association with effective radiomics dominance internally ($\rho=0.290$, 95\% CI 0.041--0.516). The external correlation estimate was smaller and remained uncertain ($\rho=0.086$, 95\% CI $-0.220$ to 0.400; Fig.~\ref{fig:structural_effective}B).

Structural weighting and effective contribution describe different stages of the model computation. Gate coefficients scale intermediate representations, whereas the branch terms additionally incorporate feature values, classifier weights, and cancellation across dimensions. Gate weighting alone is therefore insufficient to infer output-level branch contribution. At the same time, a small radiomics branch term does not imply that the classifier is independent of radiomics because radiomics also determines the gate applied to the image branch.

The distinction observed here is consistent with previous work showing
that learned weighting coefficients alone may incompletely characterise
downstream model outputs. Kobayashi et al.\ demonstrated that incorporating
the norms of transformed representations can change the interpretation
obtained from attention weights alone \cite{kobayashi2020attention}.
Although attention and gated feature fusion are different mechanisms,
both motivate examining weighted representations rather than coefficients
in isolation. Here, the analysis additionally incorporates the final
classifier weights to quantify signed contributions to the binary
decision logit. These branch terms depend on projected feature values,
gate coefficients, classifier weights, and cancellation across dimensions.
Moreover, radiomics influences the gated image term through $\mathbf{g}$,
so a small direct radiomics contribution does not imply limited total
radiomics influence. Gate statistics and classifier-level branch terms
therefore provide complementary, rather than interchangeable,
descriptions of fusion behaviour.

\subsection{Directional Agreement and Conflict}

Conflict occurred in 28.3\% of internal cases and 34.1\% of external cases (Table~\ref{tab:conflict}). Internally, conflict cases showed a numerically higher error rate than agreement cases, but neither error rates nor the distribution of $D_{\mathrm{rad}}$ differed significantly between groups (Fisher's exact $p=0.530$; Mann--Whitney $p=0.682$). External comparisons likewise showed no clear differences (Fisher's exact $p=1.000$; Mann--Whitney $p=0.967$).

These exploratory comparisons show that the two signed branch terms can reinforce or oppose one another within individual predictions. However, the results provide limited evidence that contribution conflict identifies difficult cases or that radiomics plays a corrective role. Establishing such a role would require a direct intervention on the radiomics-mediated pathway rather than comparison of signed branch terms alone.

\begin{table}[!htbp]
\centering
\caption{Directional agreement and conflict. Errors refer to the full classifier at the fixed validation-selected threshold. Mean and median effective radiomics dominance are reported for each group.}
\label{tab:conflict}
\small
\setlength{\tabcolsep}{4pt}
\begin{tabular}{llrccc}
\hline
Cohort & Relation & $n$ & Errors/$n$ (\%) & Mean $D_{\mathrm{rad}}$ & Median $D_{\mathrm{rad}}$ \\
\hline
Internal & Agreement & 43 & 11/43 (25.6) & 0.128 & 0.086 \\
 & Conflict & 17 & 6/17 (35.3) & 0.198 & 0.043 \\
\hline
External & Agreement & 27 & 7/27 (25.9) & 0.060 & 0.035 \\
 & Conflict & 14 & 4/14 (28.6) & 0.112 & 0.039 \\
\hline
\end{tabular}
\end{table}

\FloatBarrier

\subsection{Radiomics Feature Attribution}

\begin{figure}[!t]
\centering
\includegraphics[width=\textwidth]{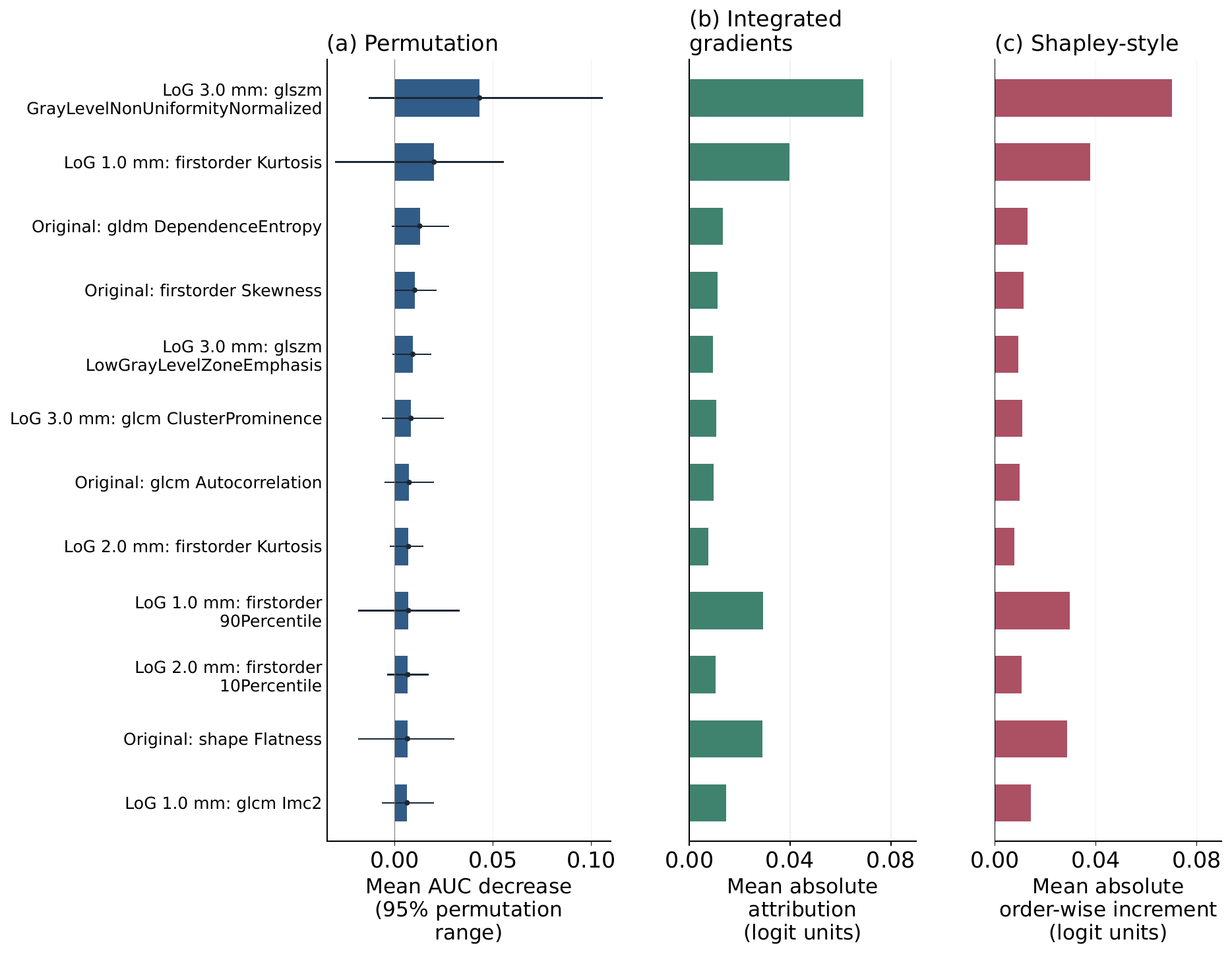}
\caption{Internal feature attribution to $C_{\mathrm{rad}}$. The 12 leading features by mean permutation-AUC decrease appear in a common row order. (a) Mean AUC decreases and central 95\% ranges across 200 within-cohort permutations; ranges describe shuffle variability rather than patient-sampling CIs. (b) Mean absolute integrated gradients. (c) Mean absolute order-wise marginal increments, averaged over 64 feature-addition orders and patients. Panels (b) and (c) are in logit units. LoG scales are in mm.}
\label{fig:attribution}
\end{figure}

The largest internal mean permutation-AUC decrease for $C_{\mathrm{rad}}$ was associated with LoG ($\sigma=3$ mm) GLSZM GrayLevelNonUniformityNormalized ($\Delta$AUC 0.043, $\Delta$AP 0.041). The same feature ranked first under both integrated-gradient and Shapley-style attribution, showing consistent prominence across the three complementary attribution analyses (Fig.~\ref{fig:attribution}).

The external permutation ranking differed, with LoG ($\sigma=1$ mm) first-order Skewness showing the largest mean AUC decrease. Given the small number of external non-clear cell RCC cases, this difference should be interpreted descriptively rather than as evidence of a reproducible shift in feature importance.

The three attribution methods characterise different properties of $C_{\mathrm{rad}}$. Permutation importance measures changes in branch-score discrimination, whereas integrated gradients and the Shapley-style analysis quantify reference-relative changes in the branch logit. Since perturbing radiomics also changes the gate applied to the image representation, these rankings should be interpreted as sensitivity of the radiomics branch term rather than importance for the complete classifier.

\subsection{Scope and Limitations}

This study characterises one fixed classifier and one internal split. The decomposition is exact for its additive fusion and linear head, while its interpretation is conditional on the chosen branch representation. The dominance ratio concerns net non-bias terms: cancellation within each branch occurs before taking absolute values, and direction is assessed separately. It is therefore a branch-level description rather than a causal allocation of information sources.

The small, positive-enriched external cohort limits generalisation. Resampling only two observed non-clear cell RCC cases cannot represent population heterogeneity, and the external association estimate remains uncertain. High external AP must also be interpreted relative to the high ccRCC prevalence. Differences in case mix and segmentation could affect transportability, but their individual effects were not isolated. Group comparisons were unadjusted, and attribution estimates depend on the chosen perturbations and reference. Unconditional shuffling and reference-based feature combinations may depart from the joint distribution of correlated radiomics measurements. These analyses support computational interpretability; biological interpretation and clinical utility would require additional validation.

\section{Conclusion}

This study shows that gate inspection and decision-level decomposition answer different interpretability questions in gated radiomics--image feature fusion. In the analysed RCC classifier, strong structural weighting toward radiomics coexisted with a substantially smaller radiomics branch contribution at the classifier output, while the gated image term retained most of the observed discriminative ranking information. Because the gated image term remains conditioned on radiomics, these findings should not be interpreted as evidence that the radiomics input is dispensable. Instead, they distinguish structural weighting, additive branch contribution, and total input dependence as related but non-equivalent descriptions of model behaviour. Reporting gate statistics together with classifier-weighted branch terms therefore provides a more precise account of gated-fusion behaviour. Future work should examine whether this structural-to-effective discrepancy persists across other fusion architectures and use intervention-based analyses to quantify total input dependence in larger and more balanced external cohorts.

\bibliographystyle{splncs04}
\bibliography{bib}

@article{arevalo2017gated,
  title={Gated multimodal units for information fusion},
  author={Arevalo, John and Solorio, Thamar and Montes-y-G{\'o}mez, Manuel and Gonz{\'a}lez, Fabio A},
  journal={arXiv preprint arXiv:1702.01992},
  year={2017}
}

@inproceedings{jain2019attention,
  title={Attention is not explanation},
  author={Jain, Sarthak and Wallace, Byron C},
  booktitle={Proceedings of the 2019 Conference of the North American Chapter of the Association for Computational Linguistics: Human Language Technologies, Volume 1 (Long and Short Papers)},
  pages={3543--3556},
  year={2019}
}

@inproceedings{wiegreffe2019attention,
  title={Attention is not not explanation},
  author={Wiegreffe, Sarah and Pinter, Yuval},
  booktitle={Proceedings of the 2019 conference on empirical methods in natural language processing and the 9th international joint conference on natural language processing (EMNLP-IJCNLP)},
  pages={11--20},
  year={2019}
}

@article{aerts2014decoding,
  title={Decoding tumour phenotype by noninvasive imaging using a quantitative radiomics approach},
  author={Aerts, Hugo JWL and Velazquez, Emmanuel Rios and Leijenaar, Ralph TH and Parmar, Chintan and Grossmann, Patrick and Carvalho, Sara and Bussink, Johan and Monshouwer, Ren{\'e} and Haibe-Kains, Benjamin and Rietveld, Derek and others},
  journal={Nature communications},
  volume={5},
  number={1},
  pages={4006},
  year={2014},
  publisher={Nature Publishing Group UK London}
}

@InProceedings{pmlr-v301-varma26a,
  title = 	 {MedVAE: Efficient Automated Interpretation of Medical Images with Large-Scale Generalizable Autoencoders},
  author =       {Varma, Maya and Kumar, Ashwin and van der Sluijs, Rogier and Ostmeier, Sophie and Blankemeier, Louis and Chambon, Pierre Joseph Marcel and Bluethgen, Christian and Prince, Jip and Langlotz, Curtis and Chaudhari, Akshay S},
  booktitle = 	 {Proceedings of The 8th International Conference on Medical Imaging with Deep Learning},
  pages = 	 {1597--1626},
  year = 	 {2026},
  editor = 	 {Tasdizen, Tolga and Elhabian, Shireen and Summers, Ronald and Chen, Chen and Koch, Lisa and Zhuang, Yan},
  volume = 	 {301},
  series = 	 {Proceedings of Machine Learning Research},
  month = 	 {09--11 Jul},
  publisher =    {PMLR},
  url = 	 {https://proceedings.mlr.press/v301/varma26a.html}
}

@misc{bex2025european,
  author  = {Bex, Axel and Abu Ghanem, Yasmin and Albiges, Laurence and others},
  title   = {European Association of Urology Guidelines on Renal Cell Carcinoma: The 2025 Update},
  journal = {European Urology},
  volume  = {87},
  number  = {6},
  pages   = {683--696},
  year    = {2025},
  doi     = {10.1016/j.eururo.2025.02.020}
}

@article{wang2021radiomics,
  title={Radiomics models based on enhanced computed tomography to distinguish clear cell from non-clear cell renal cell carcinomas},
  author={Wang, Ping and Pei, Xu and Yin, Xiao-Ping and Ren, Jia-Liang and Wang, Yun and Ma, Lu-Yao and Du, Xiao-Guang and Gao, Bu-Lang},
  journal={Scientific Reports},
  volume={11},
  number={1},
  pages={13729},
  year={2021},
  publisher={Nature Publishing Group UK London}
}

@article{borys2023explainable,
  title={Explainable AI in medical imaging: An overview for clinical practitioners--Beyond saliency-based XAI approaches},
  author={Borys, Katarzyna and Schmitt, Yasmin Alyssa and Nauta, Meike and Seifert, Christin and Kr{\"a}mer, Nicole and Friedrich, Christoph M and Nensa, Felix},
  journal={European journal of radiology},
  volume={162},
  pages={110786},
  year={2023},
  publisher={Elsevier}
}

@article{warfield2004simultaneous,
  title={Simultaneous truth and performance level estimation (STAPLE): an algorithm for the validation of image segmentation},
  author={Warfield, Simon K and Zou, Kelly H and Wells, William M},
  journal={IEEE transactions on medical imaging},
  volume={23},
  number={7},
  pages={903--921},
  year={2004},
  publisher={IEEE}
}

@article{murugesan2024ai,
  title={AI-generated annotations dataset for diverse cancer radiology collections in NCI image data commons},
  author={Murugesan, Gowtham Krishnan and McCrumb, Diana and Aboian, Mariam and Verma, Tej and Soni, Rahul and Memon, Fatima and Farahani, Keyvan and Pei, Linmin and Wagner, Ulrike and Fedorov, Andrey Y and others},
  journal={Scientific data},
  volume={11},
  number={1},
  pages={1165},
  year={2024},
  publisher={Nature Publishing Group UK London}
}

@dataset{vanoss2023aimi,
  author    = {Van Oss, Jeff and Murugesan, Gowtham Krishnan and McCrumb, Diana and Soni, Rahul},
  title     = {Image segmentations produced by BAMF under the AIMI Annotations initiative},
  year      = {2023},
  publisher = {Zenodo},
  doi       = {10.5281/zenodo.10081112}
}

@article{van2017computational,
  title={Computational radiomics system to decode the radiographic phenotype},
  author={Van Griethuysen, Joost JM and Fedorov, Andriy and Parmar, Chintan and Hosny, Ahmed and Aucoin, Nicole and Narayan, Vivek and Beets-Tan, Regina GH and Fillion-Robin, Jean-Christophe and Pieper, Steve and Aerts, Hugo JWL},
  journal={Cancer research},
  volume={77},
  number={21},
  pages={e104},
  year={2017}
}

@article{zwanenburg2020image,
  title={The image biomarker standardization initiative: standardized quantitative radiomics for high-throughput image-based phenotyping},
  author={Zwanenburg, Alex and Valli{\`e}res, Martin and Abdalah, Mahmoud A and Aerts, Hugo JWL and Andrearczyk, Vincent and Apte, Aditya and Ashrafinia, Saeed and Bakas, Spyridon and Beukinga, Roelof J and Boellaard, Ronald and others},
  journal={Radiology},
  volume={295},
  number={2},
  pages={328--338},
  year={2020},
  publisher={Radiological Society of North America}
}

@inproceedings{sundararajan2017axiomatic,
  title={Axiomatic attribution for deep networks},
  author={Sundararajan, Mukund and Taly, Ankur and Yan, Qiqi},
  booktitle={International conference on machine learning},
  pages={3319--3328},
  year={2017},
  organization={PMLR}
}

@incollection{shapley1953value,
  author    = {Shapley, Lloyd S.},
  title     = {A Value for n-Person Games},
  booktitle = {Contributions to the Theory of Games II},
  editor    = {Kuhn, Harold W. and Tucker, Albert W.},
  pages     = {307--317},
  year      = {1953},
  publisher = {Princeton University Press},
  address   = {Princeton},
  doi = {10.1515/9781400881970-018}
}

@dataset{heller2023kits23,
  author    = {Heller, Nicholas and Isensee, Fabian and Tejpaul, Resha and Wood, Andrew W. and Papanikolopoulos, Nikolaos P. and Weight, Christopher},
  title     = {2023 Kidney and Kidney Tumor Segmentation Challenge},
  year      = {2023},
  publisher = {Zenodo},
  doi       = {10.5281/zenodo.7840134}
}

@inproceedings{sun2025borrowing,
  author = {Sun, Minghui and Engelhard, Matthew M. and Goldstein, Benjamin},
  title = {Borrowing From the Future: Enhancing Early Risk Assessment through Contrastive Learning},
  booktitle = {Proceedings of the 10th Machine Learning for Healthcare Conference},
  series = {Proceedings of Machine Learning Research},
  volume = {298},
  publisher = {PMLR},
  year = {2025},
  url = {https://proceedings.mlr.press/v298/sun25a.html}
}

@inproceedings{kobayashi2020attention,
  author = {Kobayashi, Goro and Kuribayashi, Tatsuki and Yokoi, Sho and Inui, Kentaro},
  title = {Attention is Not Only a Weight: Analyzing Transformers with Vector Norms},
  booktitle = {Proceedings of the 2020 Conference on Empirical Methods in Natural Language Processing (EMNLP)},
  pages = {7057--7075},
  publisher = {Association for Computational Linguistics},
  year = {2020},
  doi = {10.18653/v1/2020.emnlp-main.574}
}

@article{shrout1979intraclass,
  author = {Shrout, Patrick E. and Fleiss, Joseph L.},
  title = {Intraclass correlations: Uses in assessing rater reliability},
  journal = {Psychological Bulletin},
  volume = {86},
  number = {2},
  pages = {420--428},
  year = {1979},
  doi = {10.1037/0033-2909.86.2.420}
}

@article{fisher2019mcr,
  author = {Fisher, Aaron and Rudin, Cynthia and Dominici, Francesca},
  title = {All Models are Wrong, but Many are Useful: Learning a Variable's Importance by Studying an Entire Class of Prediction Models Simultaneously},
  journal = {Journal of Machine Learning Research},
  volume = {20},
  number = {177},
  pages = {1--81},
  year = {2019},
  url = {https://jmlr.org/papers/v20/18-760.html}
}

@article{youden1950index,
  author = {Youden, William J.},
  title = {Index for rating diagnostic tests},
  journal = {Cancer},
  volume = {3},
  number = {1},
  pages = {32--35},
  year = {1950},
  doi = {10.1002/1097-0142(1950)3:1<32::AID-CNCR2820030106>3.0.CO;2-3}
}

@article{efron1979bootstrap,
  author = {Efron, Bradley},
  title = {Bootstrap methods: Another look at the jackknife},
  journal = {The Annals of Statistics},
  volume = {7},
  number = {1},
  pages = {1--26},
  year = {1979},
  doi = {10.1214/aos/1176344552}
}

@article{castro2009sampling,
  author = {Castro, Javier and G{\'o}mez, Daniel and Tejada, Juan},
  title = {Polynomial calculation of the {Shapley} value based on sampling},
  journal = {Computers \& Operations Research},
  volume = {36},
  number = {5},
  pages = {1726--1730},
  year = {2009},
  doi = {10.1016/j.cor.2008.04.004}
}
\end{document}